\documentclass[prl,twocolumn,showpacs,preprintnumbers,amsmath,amssymb]{revtex4}
\usepackage{graphicx}
\usepackage{dcolumn}
\usepackage{amsmath,amssymb} 
 
\newcommand{\beq}{\begin{equation}} 
\newcommand{\eeq}{\end{equation}}
\newcommand{\beqa}{\begin{eqnarray}}
\newcommand{\eeqa}{\end{eqnarray}}

\newcommand{\ad}{\gamma}	

\newcommand{\percent}{\%}
\begin{document}
\title{Pairing and polarization in systems with retarded interactions}

\author{S. Ciuchi$^1$}
\author{G. Sangiovanni$^2$}
\author{M. Capone$^{2,3}$}

\affiliation{$^1$Istituto Nazionale di Fisica della Materia and 
Dipartimento di Fisica\\
Universit\`a dell'Aquila,
via Vetoio, I-67010 Coppito-L'Aquila, Italy}
\affiliation{$^2$Istituto Nazionale di Fisica della Materia, 
Unit\`a Roma 1 and SMC Center, and Dipartimento di Fisica\\
Universit\`a di Roma "La Sapienza" piazzale Aldo Moro 5, I-00185 Roma, Italy}
\affiliation{$^3$Istituto dei Sistemi Complessi del CNR, Via dei Taurini 19, I-00185 Roma, Italy}
\pacs{71.38.-k,  71.30.+h,  71.10.Fd} 
%
\date{\today}
\begin{abstract}
In a system where a boson (e.g, a phonon) of finite frequency  $\omega_0$ is
coupled to electrons, two phenomena occur as the coupling is increased:
electron pairing  and polarization of the  boson field. Within a
path integral formalism and a Dynamical Mean-Field approach,  we introduce {\it
ad hoc} distribution  function which allow us to pinpoint the two effects. When
$\omega_0$ is smaller than the bandwidth $D$, pairing and polarization
occur for fairly similar couplings for all considered temperatures.  When
$\omega_0 > D$, the two phenomena tend to coincide only for $T \gg \omega_0$,
but are no longer tied for low  temperatures so that a state of paired
particles without finite  polarization is stabilized. 
\end{abstract} 
\maketitle

Boson-mediated attractive interactions are known to drive instabilities
toward broken symmetry phases like charge ordering and superconductivity
in electronic systems.
The crossover from weak to strong coupling  has been extensively studied in both 
superconducting \cite{Micnas} and charge ordered phases \cite{Micnas,cdp},
uncovering a general weak-to-strong coupling scenario in which
at weak coupling pairing and condensation occur at the same 
temperature scale while at strong coupling pairing occurs at a 
higher temperature than condensation, leading to an intermediate phase
of incoherent pairs.

The same evolution from weak to strong coupling in the normal phase has been
studied mainly in the limit of instantaneous attraction (attractive Hubbard
model), where the above pairs have been proposed to describe the
pseudogap regime of the cuprates. This issue, relevant also for other field of
applications such as the negative $U$ localizing centers \cite{anderson75}, is 
in its nature quite hard to attack.  In fact, the intermediate regime, in which
the system changes its nature from a Fermi liquid  to a paired state,
intrinsically requires non-perturbative approaches.

The Dynamical Mean Field Theory (DMFT) emerged in the last decade as one of the 
most successful method to deal with non-perturbative regimes.
In analogy with classical mean-field, DMFT obtains an exact solution for
local quantum dynamics at the expense of spatial fluctuations which are frozen,
and becomes exact for large coordination.
\cite{DMFTreview}.
DMFT has been successfully applied to the study of the Mott-Hubbard
transition in the repulsive Hubbard model, in which by increasing the Coulomb
repulsion, a metal becomes a Mott insulator formed by a collection of localized spins.
Since the half-filled repulsive and attractive Hubbard model can be mapped
one onto the other by a unitary particle-hole transformation on one spin
species,\cite{Micnas} 
the Mott-Hubbard transition maps onto an analogous metal-insulator transition 
of the attractive model, where the insulating phase is made of local pairs and 
empty sites.
This means that much of the knowledge we have on the Mott transition can 
be used for the study of the pairing transition of the attractive model
\cite{massimo11,metzneruneg}.
More precisely, we know that the transition is a first order line in the 
interaction-temperature diagram, ending in a finite temperature critical point 
\cite{DMFTreview}.
For temperatures larger than the critical temperature, a crossover between two
phases with different properties survives. 

This paper is mainly devoted to discuss how such a scenario in enriched 
when the attraction (or equivalently the repulsion) has a finite intrinsic 
energy scale, given by the bare frequency of the boson which mediates
the attraction. The pairing process, which becomes a bipolaronic transition
when the mediator boson is a phonon, survives the inclusion of boson 
dynamics, and it has been studied within DMFT by various
authors \cite{massimo4,bulla}.
The new energy scale given by the intrinsic frequency for the boson dynamics
has to be compared with both the temperature and the electron half-bandwidth $D$.
The most interesting region, in which the use of DMFT is crucial, is the 
intermediate region in which none of the energy scale and the temperature
are negligible.
The ratio  $\gamma=\omega_0/D$ between the boson energy and the electron
hopping energy naturally defines an adiabatic regime in which $\gamma$ is small 
and the electronic degrees of freedom are much faster than bosonic ones. 
In this regime a gradual increase of the coupling determines a crossover from 
quasi-free electrons to almost localized ``polarons'', quasiparticles
in which the electrons are strongly bound to the bosonic degrees of 
freedom thereby determining a polarization of the boson field which couples
to the electron density. Such a crossover can be unambiguously defined
by looking at the polarization properties of the boson field
\cite{freericks1,millisI}. 
In the opposite antiadiabatic limit, in which the boson dynamics is faster
than the electronic one, the interaction becomes instantaneous, and we
recover either the attractive or repulsive Hubbard models.

The main result of this work is to clearly disentangle the
 polarization and the pairing crossovers in the normal phase
and to define unambiguously also the pairing crossover.
We find that, the larger is $gamma$, the more the two processes occur at different coupling, 
opening an interesting regime in between.

Although we do not need to specify the physical origin of the bosonic mediator,
we consider for the sake of definiteness the Holstein model which  describes
tight-binding electrons coupled to
dispersionless Einstein phonons.  Within DMFT the lattice Holstein model  is
mapped onto a local impurity problem whose action reads\cite{freericks1}:
\begin{equation}\label{Sel}
S_{el} = -\int_0^\beta d\tau \int_0^\beta d\tau' \sum_{\sigma} c_{\sigma}^\dagger(\tau) \mathcal{G}^{-1}_0(\tau-\tau') c_{\sigma}(\tau') 
\end{equation}
\begin{equation}\label{Sph}
S_{ph} = \frac{1}{2} \int_0^\beta d \tau \left( \frac{\dot{x}^2(\tau)}{\omega_0^2} + x^2(\tau) \right) 
\end{equation}
\begin{equation}\label{Selph}
S_{el-ph} = - \sqrt{U} \int_0^\beta d\tau x(\tau) \left(n(\tau) - 1 \right) 
\end{equation}
where $n(\tau) = \sum_{\sigma} c_{\sigma}^\dagger(\tau) c_{\sigma}(\tau)$. We
take  $\kappa = m \omega_0^2 = 1$,  $m$ being the mass of the oscillator.
In Eq. (\ref{Sel}) $\mathcal{G}^{-1}_0(\tau)$ is the bath propagator that
has to  be evaluated self-consistently \cite{DMFTreview}. 
The self-consistency condition which enforces the DMFT contains the 
informations about the original lattice, and for 
 the infinite coordination Bethe lattice we consider it takes the simple form
$\mathcal{G}^{-1}_0(\tau)=-\partial_\tau+(D^2/4) G(\tau)$ where $G(\tau)$ is the 
local Green's function and $D$ is the
half-bandwidth. For our model with dispersionless Einstein modes, the phononic 
degrees of freedom are not subject to an analogous self-consistency.
Since the action is manifestly particle-hole symmetric, it describes
a half-filled system with density $n=1$.
In the antiadiabatic limit ($\omega_0=\infty$, $U$ finite) the kinetic term of
the boson field vanishes, the interaction becomes therefore instantaneous 
and the action coincides with that of an attractive  Hubbard
model after an  Hubbard-Stratonovich (HS) decoupling of the
quartic interaction term. 
We notice incidentally that under the transformation which maps
the attractive Hubbard model on the repulsive one, (\ref{Selph}) becomes
$S_{el-ph} = - \sqrt{U} \int_0^\beta d\tau x(\tau) \left(n_{\uparrow}(\tau) -
n_{\downarrow}(\tau)\right) $
which, once the spin index is interpreted as an orbital one,
has been studied in Ref. \onlinecite{pata2} to describe
the Jahn-Teller (JT) coupling in manganites. 

An efficient method to solve (\ref{Sel},\ref{Sph},\ref{Selph}) at finite temperature is 
the Quantum Monte Carlo method (QMC) in the Blankenbecler-Scalapino-Sugar (BSS)
approach\cite{BSS}. This method works well in the range of temperatures we are
interested in (i.e. not too low temperatures) and naturally yields
electronic and bosonic correlation
functions, as well as the probability distributions associated to the boson
fields. The method is not affected by the negative
sign-problem, and its main limitation comes in the adiabatic regime
($\gamma \ll 1$) where the boson becomes heavy making 
more difficult to sample correctly the available phase space. 
In the BSS scheme the fermions
are integrated out , and the bosons coordinates $x(\tau)$ are
discretized into $L$ imaginary-time slices of width $\Delta \tau = \beta/L$ and
then sampled by the Monte Carlo simulation.
$L$ has to be chosen large enough to reduce as much as possible
$\Delta\tau$, which  controls the the Trotter discretization error.
To keep $\Delta \tau$
less than $1/8$ we used  $32$ slices except for the 
lowest temperature ($\beta=8$)
for which we have used $L=64$,
As customary,  the statistical error is reduced by dividing
measurements in bins.
The self-consistency is achieved in a few DMFT iterations, typically from $5$ 
to $10$, depending on the values of the parameters\cite{DMFTreview}.

Now we introduce the key quantities we employ for the characterization 
of pairing and polarization and their temperature dependence.
Namely, we compute  the probability distributions of the endpoint
\beq
\label{PX}
P(X)= \left < \delta(X-x(0)) \right >,
\eeq
and of the center of mass $X_c$ (``centroid'') of the boson path in imaginary time
\beq
\label{PXc}
P(X_c)= \left < \delta(X_c-\frac{1}{\beta}\int_0^\beta x(\tau)d\tau) \right >
\eeq
where the averages are evaluated over the action given by Eqs. (\ref{Sel}-\ref{Selph}). 

The first quantity has a rather straightforward meaning as a probability 
distribution of the lattice displacement (if the boson is thought as a phonon).
If $P(X)$ has a single maximum, corresponding to a uniform displacement of
the field, the system is not polarized. 
A lattice polarization reflects in 
the presence of two maxima in $P(X)$, 
corresponding to opposite polarization of
occupied and unoccupied sites (bimodal behavior) \cite{millisI}.
In this way a {\it qualitative} difference is identified between the 
polarized and unpolarized regimes, which allows for an unambiguous way
to draw a crossover line, as opposed to estimates based on smoothly
varying functions as average lattice fluctuations
or electron kinetic energy.

The meaning of the centroid variable $X_C$ has been discussed in 
\cite{Kleinert} for a single particle in a binding potential.
Here the variable $X$ represents the position of the particle, and
 $X_c$ is the classical position of the  particle \cite{Kleinert}. 
For an heavy particle, the classical limit holds, so that   (\ref{PX}) and (\ref{PXc}) 
coincide \cite{Kleinert}.
As the particle is made lighter, the wave function becomes broader, increasing
the variance of $P(X)$ while $P(X_c)$ turns out to be essentially determined by
the binding range of the potential.
Here, we use $P(X_C$) for the many-body problem, and propose that pairing, i.e., 
the formation of local pairs, can be associated with a 
development of a multimodal behavior in our centroid distribution
$P(X_c)$. This estimator of the pairing crossover has the same advantage
of the previous one, and it finally allows us to  draw, beside the polarization
line $U_{pol}$, an equally unambiguous pairing line $U_{pair}(T)$ for
any value of the boson frequency.
A crossover line can also be associated with  maxima of
susceptibilities related to the relevant quantity,  e.g., the
electron-phonon  correlation function in the case of polaron crossover
\cite{ciukIntJourn}. However, this choice is not obvious in  the case of pairing,
where charge, superconductivity or Pauli \cite{sewer}  susceptibilities can
play this role. We defer a comparison with these methods in the concluding
remarks.

The ability of our estimator to determine the pairing crossover can be 
understood by considering directly the interaction term (\ref{Selph}) in the
action.
In the adiabatic limit ($\ad \rightarrow 0$) the kinetic term 
forces the boson path to be $\tau$-independent. The boson field becomes 
classical and 
the interaction term  can be written as $-\sqrt{U} X \int_0^\beta
d\tau (n(\tau)-1)$, where $X$ indicates the constant value assumed by
$x(\tau)$ along the whole imaginary time path. In this limit the centroid 
coordinate $X_c$ is equal to $X$ and the two distributions $P(X)$ and $P(X_c)$ 
obviously coincide. Thus the centroid distribution becomes bimodal when the system
is polarized, which is exactly what one expects in the static limit, 
since a static field can induce pairing only with a finite polarization.

On the other hand, in the opposite atomic ($D \rightarrow 0$, 
$\ad \rightarrow \infty $)
limit the electron density becomes a constant of motion. 
Therefore Eq. (\ref{Selph}) takes the transparent form 
$-\sqrt{U} (n-1) \int_0^\beta d\tau x(\tau)$ where the electron density is 
directly coupled to the centroid $X_c$. 
The average appearing in Eq. (\ref{PXc}) is readily carried out, giving
\beq
\label{PXcAtomic}
P(X_c) \propto \exp \left[-\beta (\frac{X_c^2}{2}-\frac{1}{\beta}\log (2
\cosh (\beta\sqrt{U}X_c+1))) \right]
\eeq
which becomes bimodal for $T< U/2$. This is exactly the scale where double
occupancies start to proliferate in the atomic limit. 
Therefore  the bimodality of $P(X_c)$ 
correctly signals the onset of pairing also in the antiadiabatic regime. 
In the same limit, it can be proved that the endpoint distribution $P(X)$ has
a variance which scales with $1/\sqrt{\Delta \tau}$ 
and as a consequence no definite polarization may occur.
We finally notice that the $D \rightarrow 0$ 
limit of adiabatic $P(X)$ \cite{millisI}
coincides with $P(X_c)$ of Eq. (\ref{PXcAtomic}).
Since for  $\omega_0=0$ the distributions of $X$ and $X_c$ coincide, 
we conclude that in the atomic limit $P(X_c)$ 
is the same for $\omega_0=0$ and 
$\omega_0=\infty$. This suggests that the pairing crossover may depend
on $\omega_0$ more weakly than the polarization one.

\begin{figure}[htbp]
\begin{center}
\includegraphics[width=6.5cm,angle=270]{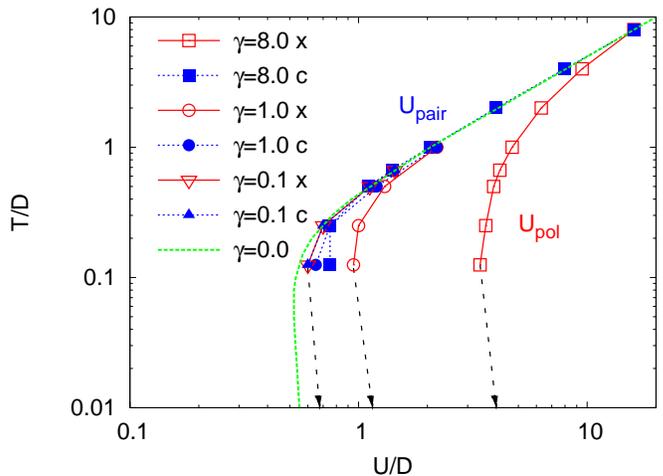}
\end{center}
\caption{(color online) Behavior of $U_{pair}(T)$ (in blue and 
with the letter $c$) and $U_{pol}(T)$ (in red and with the letter 
$x$) at $\ad=0.1$, $1$ and $8$, respectively indicated by solid 
and open triangles, solid and open circles and solid and open squares. 
The green dashed line represents the $\ad = 0$ result for both 
$U_{pair}(T)$ and $U_{pol}(T)$. The dashed arrows indicates 
the zero-temperature results for the polaron crossover \cite{massimo4}.}
\label{PD}
\end{figure}

To analyze the evolution of $P(X)$ and $P(X_c)$ at finite $D$ and $\omega_0$ 
we use BSS-QMC. The numerically exact results, shown in Fig.\ref{PD}, 
clearly show that $P(X)$ and $P(X_c)$ tend to coincide in the 
relatively adiabatic case $\ad = 0.1$, as expected from the previous
arguments about the adiabatic limit. The two quantities are clearly 
 different for $\ad = 1$ and $8$. 
Actually for temperatures smaller than $\omega_0$, the polarization
crossover $U_{pol}$ moves to larger
values as $\ad $ is increased, while  the line ($U_{pair}$) where $P(X_c)$
becomes bimodal is only slightly shifted to larger couplings with increasing
$\ad$. This is strongly reminiscent of the behavior of the metal-insulator
transition in the Holstein model at $T=0$, whose critical coupling
is slowly increasing with $\ad$ and then saturates to the asymptotic
$\ad=\infty$ value\cite{massimo4}. On the other hand, the polarization
crossover is almost directly proportional to $\ad$\cite{massimo4}.
Both at zero and at finite temperature the line where the centroid becomes
bimodal does not coincide with the metal-insulator line, but it can be
considered a precursor which depends in a very similar way on $\ad$ 
and on $U/D$.

Our DMFT results can also be compared with the semi-analytical results for
$\ad=0$ \cite{millisI}, which is represented by the green dashed line 
in Fig.\ref{PD}. It can be seen that the $\ad=0.1$ case is in very good
agreement with the adiabatic result, and also the $\ad=1$ and $8$ cases,
at high temperature, fall on the same curve. It must be observed also that, in
the full DMFT calculation, the centroid distribution depends weakly on
$\ad$, as suggested by the atomic limit. 
Interestingly, the adiabatic result displays a re-entrance at low temperatures.
Although the QMC simulations do not reach sufficiently low temperatures, we
find that the re-entrance is present also for $\ad$ different from zero,
as indicated by the arrows in Fig.\ref{PD}, which mark $T=0$ results for the 
polarization crossover for the Holstein model\cite{massimo4}. 
This phenomenon has been also reported for the Mott transition in the repulsive
model \cite{chitra1}, and associated to spin entropy of the insulator.
The same physics holds here, where the entropy is associated to a 
pseudo-spin.

\begin{figure}[htbp]
\begin{center}
\includegraphics[width=6.5cm,angle=270]{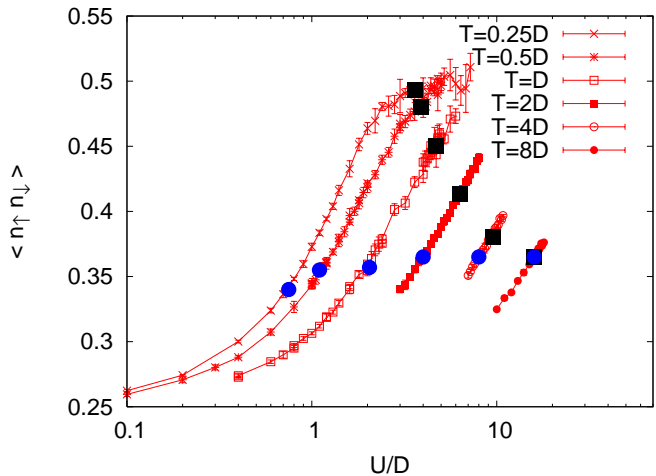}
\end{center}
\caption{(color online) Double occupancies for different temperatures at $\ad=8$. The
blue solid circle (black solid square) denotes the coupling where $P(X_c)$
($P(X)$) develops the bimodal character.}
\label{DOcc}
\end{figure}

A strong indication that the centroid distribution marks the pairing
crossover is provided by the analysis of the  double
occupancies $\langle n_\uparrow n_\downarrow \rangle$. 
This quantity increases from the non-interacting value $1/4$ to 
the asymptotic value of $1/2$ as the coupling is increased.
We computed $\langle n_\uparrow n_\downarrow \rangle$ 
for different temperatures in the antiadiabatic regime $\ad=8$, 
chosen in order to have clearly different values for $U_{pol}$ and $U_{pair}$.
We notice that our results are correctly close to the attractive 
Hubbard model.\cite{paolo}
In Fig.\ref{DOcc} $U_{pol}$ and $U_{pair}$ are marked, respectively, by a black 
solid square and a blue solid circle, for each temperature. 
$\langle n_\uparrow n_\downarrow \rangle$ at 
$U_{pol}$ is rapidly increases with decreasing temperature, while 
$\langle n_\uparrow n_\downarrow \rangle$ calculated at $U_{pair}$ is almost 
constant with temperature, or even slightly decreasing at low $T$. So,
when the centroid distribution becomes multimodal, the double occupancies are
approximately $70\percent$ of the saturation value of $1/2$.
Similar results are found also for $\ad=1$ and $0.1$, where it is
however harder to distinguish $U_{pair}$ from $U_{pol}$.
This establishes the efficiency of our centroid distribution in
pinpointing the pairing crossover for the whole range of parameters.

In conclusion for a half-filled system subject to retarded attractive 
(repulsive) interaction we observe three qualitatively different behaviors
{\it i)} a normal state of delocalized particles, 
{\it ii)} a pair region of bounded incoherent pairs  with no
associated polarization, 
{\it iii)} a  bipolaron region in which pairing is
associated  with a definite polarization.
These three regions are delimited by two clear-cut lines where the probability
distributions for the centroid and for the endpoint of the bosonic path,
develop a bimodal shape. While the border between the normal and the local pair
phase is weakly dependent on the adiabatic ratio, the one
between the pair and the bipolaronic phase, shifts to higher value of the
coupling with increasing
$\ad$, so that the  pair phase gets larger, eventually covering
the whole right hand part of the phase diagram in the attractive Hubbard limit
($\ad = \infty$). The $U_{pair}$ line is very similar to the $T^*$ line
found in Ref.\cite{sewer} by looking at the maximum of the spin susceptibility.
For temperatures of the order of $\omega_0$ the two lines merge because phonons
at such temperatures behaves classically. 
Finally, due to the electron-hole symmetry the retarded attractive interaction
is mapped onto a JT model leading to an Hubbard-like repulsion in the 
antiadiabatic limit. In this case the
crossover at $U_{pair}$ is associated to the formation of local orbital ordering for JT
interactions. The crossover at $U_{pol}$ survives only in the retarded JT case
and is shifted toward higher value of the coupling as the phonon frequency
increase leaving a region in which orbital ordering is not associated with
JT polarons.

We acknowledge useful discussions with C. Castellani and A. Toschi.
This work was supported by MIUR-Cofin 2003 matching funds
programs.


\end{document}